\documentstyle[amssymb,twocolumn,aps]{revtex}

\begin{document}
\title{Precision calculation of magnetization and specific heat of vortex liquids
and solids in type II superconductors}
\author{Dingping Li and Baruch Rosenstein}
\address{{\it National Center for Theoretical Sciences and} \\
{\it Electrophysics Department, National Chiao Tung University } \\
{\it Hsinchu 30050, Taiwan, R. O. C.}}
\date{\today}
\maketitle

\begin{abstract}
A new systematic calculation of magnetization and specific heat
contributions of vortex liquids and solids (not very close to the melting
line) is presented. We develop an optimized perturbation theory for the
Ginzburg - Landau description of thermal fluctuations effects in the vortex
liquids. The expansion is convergent in contrast to the conventional high
temperature expansion which is asymptotic. In the solid phase we calculate
first two orders which are already quite accurate. The results are in good
agreement with existing Monte Carlo simulations and experiments. Limitations
of various nonperturbative and phenomenological approaches are noted. In
particular we show that there is no exact intersection point of the
magnetization curves both in 2D and 3D.
\end{abstract}




It was clearly seen in both magnetization \cite{Zeldov} and specific heat
experiments \cite{Schilling} that thermal fluctuations in high $T_{c}$
superconductors are strong enough to melt the vortex lattice into liquid
over large portions of the phase diagram. The transition line between the
Abrikosov vortex lattice and the liquid is located far below the mean field
phase transition line. Between the mean field transition line and the
melting point physical quantities like the magnetization, conductivity and
specific heat depend strongly on fluctuations. Several experimental
observations call for a refined precise theory. For example, a striking
feature of magnetization curves intersecting at the same point $(T^{\ast
},H^{\ast })$ was observed in a wide rage of magnetic fields in both the
layered \cite{Magn2D} materials and the more isotropic ones \cite{Magn3D}.
To develop a quantitative theory of these fluctuations, even in the case of
the lowest Landau level (LLL) corresponding to regions of the phase diagram
''close'' to $H_{c2}$\cite{Li1}, is a very nontrivial task and several
approaches were developed.

Thouless and Ruggeri \cite{Ruggeri} proposed a perturbative expansion around
a homogeneous (liquid) state in which all the ''bubble'' diagrams are
resummed. Unfortunately they proved that the series are asymptotic and
although first few terms provide accurate results at very high temperatures,
the series become inapplicable for LLL dimensionless temperature $a_{T}$ $%
\thicksim $ $(T-T_{mf}(H))/(TH)^{1/2}$ smaller than $2$ in 2D quite far
above the melting line (believed to be located around $a_{T}=-12)$.
Generally attempts to extend the theory to lower temperatures by the Borel
transform or Pade extrapolation were not successful \cite{Pade}. Several
nonperturbative methods have been also attempted including renormalization
group \cite{RG} and the $1/N$ expansion \cite{largeN}. Tesanovic and
coworkers developed a theory based on separation of the two energy scales 
\cite{Tesanovic}: the condensation energy (98\%) and the motion of the
vortices (2\%). The theory explains the intersection of the magnetization
curves.

In the first part of paper we apply optimized perturbation theory (OPT)
first developed in field theory \cite{Stevenson,Kleinert} to both the 2D and
3D LLL model. It allows to obtain a convergent (rather than asymptotic)
series for magnetization and specific heat of vortex liquids together with
precision estimate. The radius of convergence is $a_{T}=-3$ in 2D and $%
a_{T}=-5$ in 3D. On the basis of this one can make several definitive
qualitative conclusions.

Our starting point is the Ginzburg-Landau free energy: 
\begin{equation}
F=L_{c}\int d^{2}x\frac{{\hbar }^{2}}{2m}\left| {\bf D}\psi \right|
^{2}+a|\psi |^{2}+\frac{b^{\prime }}{2}|\psi |^{4},
\end{equation}
where ${\bf A}=(By,0)$ describes a nonfluctuating constant magnetic field in
Landau gauge and ${\bf D}\equiv {\bf \nabla }-i\frac{2\pi }{\Phi _{0}}{\bf A,%
}\Phi _{0}\equiv \frac{hc}{2e}$, $L_{c}$ is the width (for simplicity we
write expressions for the 2D case, essential 3D complications are discussed
separately). For simplicity we assume $a(T)=\alpha T_{c}(1-t)$, $t\equiv
T/T_{c}$. On LLL, the model after rescaling reduces to 
\begin{equation}
f=\frac{1}{4\pi }\int d^{2}x\left[ a_{T}|\psi |^{2}+\frac{1}{2}|\psi |^{4}%
\right] ,  \label{GL}
\end{equation}
where the LLL reduced temperature $a_{T}\equiv -\sqrt{\frac{4\pi }{b\omega }}%
\frac{1-t-b}{2}$ is the only parameter in the theory \cite{Ruggeri}. Here $%
b\equiv \frac{B}{H_{c2}}$, $\omega \equiv \left( 32\pi ^{3}e^{2}\kappa
^{2}\xi ^{2}T\right) /\left( c^{2}h^{2}L_{z}\right) $.

We will use a version of OPT, the optimized gaussian series \cite{Kleinert}.
It is based on the ''principle of minimal sensitivity'' \ idea \cite
{Stevenson}, first introduced in quantum mechanics. Generally a perturbation
theory starts from dividing the Hamiltonian into a solvable ''large'' part $%
K $ and a perturbation $V$. Since we can solve any quadratic Hamiltonian we
have a freedom to choose ''the best'' such quadratic part. Quite generally
such an optimization converts an asymptotic series into a convergent one
(see a comprehensive discussion, references and a proof in \cite{Kleinert}).

Due to the translational symmetry of the vortex liquid there is just one
variational parameter, $\varepsilon ,$ in the free energy divided as
follows: 
\begin{equation}
K=\frac{\varepsilon }{4\pi }|\psi |^{2},\;V=\frac{1}{4\pi }\left[ a_{H}|\psi
|^{2}+\frac{1}{2}|\psi |^{4}\right]  \label{feym}
\end{equation}
where $a_{H}\equiv a_{T}-\varepsilon $. One reads Feynman rules from eq.(\ref
{feym}): $K$ determines the propagator (just a constant), the first term in $%
V$ is a ''mass insertion'' vertex with a value of $\frac{1}{4\pi }a_{H}$ ,
while the four line vertex is $\frac{1}{8\pi }$. To calculate the effective
free energy density $\ f_{eff}=-4\pi \ln Z$, one draws all the connected
vacuum diagrams. We calculated directly diagrams up to the three loop order.
However to take advantage of the existing long series of the non optimized
gaussian expansion, we found a relation of the OPE to these series.
Originally Thouless and Ruggeri calculated these series $f_{eff}$ to sixth
order, but it was subsequently extended to $12^{th}$ ($9^{th}$ in 3D$)$ by
Brezin et al and to $13^{th}$ by Hu et al\ \cite{Hikami}. It is usually
presented using variable $x$ introduced by Thouless and Ruggeri \cite
{Ruggeri} $x=\frac{1}{\varepsilon ^{2}},\varepsilon =\frac{1}{2}\left( a_{T}+%
\sqrt{a_{T}^{2}+16}\right) $ as follows: $f_{eff}=2\log \frac{\varepsilon }{%
4\pi ^{2}}+2\sum_{n=1}^{\infty }c_{n}x^{n}.$ We can obtain all the OPT
diagrams which do not appear in the gaussian theory by insertions of bubbles
and mass insertions from the diagrams contributing to the nonoptimized
theory. Bubbles or ''cacti'' diagrams are effectively inserted by a
technique known in field theory \cite{Barnes}: 
\begin{eqnarray}
\ f_{eff} &=&2\log \frac{\varepsilon _{1}}{4\pi ^{2}}+2\sum_{n=1}^{\infty
}c_{n}x^{n} \\
x &=&\frac{\alpha }{\varepsilon _{1}^{2}},\varepsilon _{1}=\frac{1}{2}\left(
\varepsilon _{2}+\sqrt{\varepsilon _{2}^{2}+16\alpha }\right) .  \nonumber
\end{eqnarray}
Summing up all the insertions of the mass vertex is achieved by $\varepsilon
_{2}=\varepsilon +\alpha a_{H}.$ Here $\alpha $ was introduced to keep track
of order of the perturbation, so that expanding $f_{eff}$ to order $\alpha
^{n+1},$ and then taking $\alpha =1$ we obtain $\widetilde{f}%
_{n}(\varepsilon )$ (calculating $\widetilde{f}_{n}$ that way, we checked
that indeed the first three orders agree with the direct calculation). The $%
n^{th}$ OPT approximant $f_{n}$ \ is obtained by minimization of $\widetilde{%
f}_{n}(\varepsilon )$ with respect to $\varepsilon $: 
\begin{equation}
\left( \frac{\partial }{\partial \varepsilon }-\frac{\partial }{\partial
a_{H}}\right) \widetilde{f}_{n}\left( \varepsilon ,a_{H}\right) =0.
\label{mini}
\end{equation}
The above equation is equal to $1/\varepsilon ^{2n+3}$ times a polynomial $%
g_{n}\left( z\right) $ of order $n$ in $z\equiv \varepsilon \cdot a_{H}$.
That eq.(\ref{mini}) is of this type can be seen by noting that\ the
function $f$ depends on the combination $\alpha /\left( \varepsilon +\alpha
a_{H}\right) ^{2}$ only. We were unable to prove this, but have checked it
to the $40^{th}$ order. This property greatly simplifies the task: one has
to find roots of polynomials rather than solving transcendental equations.
There are $n$ (real or complex) solutions for $g_{n}\left( z\right) =0$.
However (as in the case of anharmonic oscillator \cite{Kleinert}) \ the best
results gives a real root with the smallest absolute value. We then obtain $%
\varepsilon (a_{T})=\frac{1}{2}\left( a_{T}+\sqrt{a_{T}^{2}-4z_{n}}\right) $
solving $z_{n}=\varepsilon \cdot a_{H}=\varepsilon a_{T}-\varepsilon ^{2}$.

On Fig. 1 we present OPT for different orders including $n=0$ (gaussian)
together with several orders of the nonoptimized high temperature expansion.
One observes that the OPT series converge above $a_{T}=-2.5$ and diverge
below $a_{T}=-3.5$. The proof of convergence is analogous to that for the
anharmonic oscillator, see ref. \cite{Kleinert}. On the other hand, the
nonoptimized series never converge despite the fact that above $a_{T}=2$
first few approximants provide a precise estimate consistent with OPT. Above 
$a_{T}=3$ the liquid becomes essentially a normal metal and fluctuations
effects are negligible (see Fig. 2, 3). Therefore the information the OPT
provides is essential to compare with experiments on magnetization and
specific heat. If precision is defined as $\left( f_{12}-f_{10}\right)
/f_{10}$, we obtain $4.87\%,1.27\%,0.387\%,0.222\%,$ \ $0.032\%$ at $%
a_{T}=-2,-1.5,-1,-0.5,0$ respectively. For comparison with other theories
and experiments on Fig. 2 and 3 we use the $10^{th}$ approximant.

The calculation is basically the same in 3D, the only complication being
extra integrations over momenta parallel to the magnetic field. However
since the propagator factorizes,\ these integrations can be reduced to
corresponding integrations in quantum mechanics of the anharmonic oscillator 
\cite{Ruggeri,Stevenson}. The series converge above $a_{T}=-4.5$ and diverge
below $a_{T}=-5.5.$ The nonoptimized series are useful only above $a_{T}=-1$%
. The agreement is within the expcted precision when we compare our results
in 3D with ref. \cite{Sasik}.

Now we turn to the vortex solids. Here the minimization is significantly
more difficult due to reduced symmetry. Unlike in the liquid the field $\psi 
$ acquires a nonhomogeneous expectation value and can be expressed as $\psi
(x)=v(x)+\chi (x),$where $\chi $ describes fluctuations. Assuming hexagonal
symmetry, it should be proportional to the mean field solution $%
v(x)=v\varphi _{k=0}(x)$ with a variational parameter $v$ taken real thanks
global $U(1)$ gauge symmetry where $\varphi _{k}(x)$ is the quasi - momentum
basis on LLL\cite{Li1}. Expanding $\chi $ 
\begin{equation}
\chi (x)=\frac{1}{2\pi \sqrt{2}}\int_{k}\exp [-i\theta _{k}/2]\varphi
_{k}(x)\left( O_{k}+iA_{k}\right) .  \label{expan}
\end{equation}
where real fields $A_{k}=A_{-k}^{\ast }$ $(O_{k}=O_{-k}^{\ast })$ describing
acoustic (optical) phonons of the flux lattice. The phase $\exp [-i\theta
_{k}/2]$ defined, as in the low temperature perturbation theory developed
recently \cite{Rosenstein}, via $\gamma _{k}=|\gamma _{k}|\exp [i\theta
_{k}],$ $\gamma _{k}\equiv \left\langle \varphi _{0}(x)\varphi
_{0}(x)\varphi _{k}^{\ast }(x)\varphi _{k}^{\ast }(x)\right\rangle _{x}$, is
crucial for simplification of the problem. The most general quadratic form
is 
\begin{eqnarray}
K &=&\frac{1}{8\pi }%
\int_{k}O_{k}G_{OO}^{-1}(k)O_{-k}+A_{k}G_{AA}^{-1}(k)A_{-k}+  \nonumber \\
&&O_{k}G_{OA}^{-1}(k)A_{-k}+A_{k}G_{OA}^{-1}(k)O_{-k},
\end{eqnarray}
with matrix of functions $G(k)$ to be determined together with the constant$%
\ v$ by the variational principle. The corresponding gaussian free energy $%
f_{eff}$ \ is 
\begin{eqnarray*}
&&a_{T}v^{2}+\frac{\beta _{A}}{2}v^{4}-2 \\
&&-\left\langle \log \left[ \left( 4\pi \right) ^{2}\det (G)\right]
-a_{T}\left( G_{OO}\left( k\right) +G_{AA}\left( k\right) \right)
\right\rangle _{k} \\
&&+\left\langle v^{2}\left[ \left( 2\beta _{k}+\left| \gamma _{k}\right|
\right) G_{OO}\left( k\right) +\left( 2\beta _{k}-\left| \gamma _{k}\right|
\right) G_{AA}\left( k\right) \right] \right\rangle _{k} \\
&&+\left\langle \beta _{k-l}\left[ G_{OO}\left( k\right) +G_{AA}\left(
k\right) \right] \left[ G_{OO}\left( l\right) +G_{AA}\left( l\right) \right]
\right\rangle _{k,l} \\
&&+\frac{1}{2\beta _{A}}\left\{ \left\langle \left| \gamma _{k}\right|
\left( G_{OO}\left( k\right) -G_{AA}\left( k\right) \right) \right\rangle
_{k}^{2}+4\left\langle \left| \gamma _{k}\right| G_{OA}\left( k\right)
\right\rangle _{k}^{2}\right\} 
\end{eqnarray*}
where $\left\langle ...\right\rangle _{k}$ denotes average over Brillouin
zone $\beta _{k}\equiv \left\langle \varphi _{0}^{\ast }(x)\varphi
_{0}(x)\varphi _{k}^{\ast }(x)\varphi _{k}(x)\right\rangle _{x},\beta
_{A}=\beta _{0}$. The gap equations obtained by the minimization of the free
energy look quite intractable, however they can be simplified. The crucial
observation is that $G_{OA}(k)=0$ is a solution and general solution can be
shown to differ from this simple one just by a global gauge
transformation.One can set matrix $G^{-1}$ \ as $\left( 
\begin{array}{cc}
E(k)+\Delta \left| \gamma _{k}\right|  & 0 \\ 
0 & E(k)-\Delta \left| \gamma _{k}\right| 
\end{array}
\right) $, where $\Delta $ is a constants (details will appear elsewhere).
The function $E(k)$ and the constant $\Delta $ satisfy: 
\begin{eqnarray}
E(k) &=&a_{T}+2v^{2}\beta _{k}+2\left\langle \beta _{k-l}\left( \frac{1}{%
E_{O}(l)}+\frac{1}{E_{A}(l)}\right) \right\rangle _{l}  \label{mode} \\
\beta _{A}\Delta  &=&a_{T}-2\left\langle \beta _{k}\left( \frac{1}{E_{O}(k)}+%
\frac{1}{E_{A}(k)}\right) \right\rangle _{k}.  \nonumber
\end{eqnarray}
Observing that $\beta _{k}$ has a very effective expansion in $\chi \equiv
\exp [-a_{\Delta }^{2}/2]=\exp [-2\pi /\sqrt{3}]=0.0265,\beta
_{k}=\sum_{n=0}^{\infty }\chi ^{n}\beta _{n}(k),\;\beta _{n}(k)\equiv
\sum_{\left| {\bf X}\right| ^{2}=na_{\Delta }^{2}}\exp [i{\bf k\bullet X}]$
\ and \ using the hexagonal symmetry of the spectrum, $E(k)$ can also be
expanded in ''modes'' $E(k)=\sum E_{n}\beta _{n}(k).$ The integer $n$
determines the distance of a points on the hexagonal lattice ${\bf X}$ from
the origin. One estimates that $E_{n}\simeq \chi ^{n}a_{T},$ therefore the
coefficients decrease exponentially with $n$. For some integers, for
example, $n=2,5,6$, $\beta _{n}=0$. We minimized numerically the gaussian
energy by varying $v,\Delta $ and first few modes of $E(k).$ In practice two
modes are quite enough. The results show that around $a_{T}<-5$, the
gaussian liquid energy is larger than the gaussian solid energy. So
naturally when $a_{T}<-5$, one should use the gaussian solid to set up a
perturbation theory instead of the liquid one. The gaussian energy in either
liquid (see line T0 on Fig.1) or solid\ is a rigorous upper bound on the
free energy. We calculated the leading correction (without its minimization)
in order to determine the precision of the gaussian result (see Fig. 3 for
the specific heat results). We obtain $0.2\%$, $0.4\%$ and $2\%$ at $%
a_{T}=-30,-20,-12$ respectively.

In the rest of the paper we compare our results with other theories,
simulations and experiments. An analytic theory used successfully to fit the
magnetization and the specific heat data \cite{Pierson} was developed in 
\cite{Tesanovic}. Their \ free energy density is: 
\begin{eqnarray}
f_{eff} &=&-\frac{a_{T}^{2}U^{2}}{4}+\frac{a_{T}U}{2}\sqrt{\frac{%
U^{2}a_{T}^{2}}{4}+2}+2arc\sinh \left[ \frac{a_{T}U}{2\sqrt{2}}\right]
\label{teseq} \\
U &=&\frac{1}{2}\left[ \frac{1}{\sqrt{2}}+\frac{1}{\sqrt{\beta _{A}}}+\tanh %
\left[ \frac{a_{T}}{4\sqrt{2}}+\frac{1}{2}\right] \left( \frac{1}{\sqrt{2}}-%
\frac{1}{\sqrt{\beta _{A}}}\right) \right] .  \nonumber
\end{eqnarray}
The corresponding magnetization and specific heat are shown as a dashed
lines on Fig.2 and 3 respectively. At large positive $a_{T},$ $f_{eff}=2\log
a_{T}+\frac{4}{a_{T}^{2}}-\frac{16}{a_{T}^{4}}+\frac{320}{3a_{T}^{6}}$ and
differs very little from\ the exact series $2\log a_{T}+\frac{4}{a_{T}^{2}}-%
\frac{18}{a_{T}^{4}}+\frac{1324}{9a_{T}^{6}}.$ It's low temperature
asymptotics is however less precise:$-\frac{a_{T}^{2}}{2\beta _{A}}-2\log 
\frac{\left| a_{T}\right| }{4\pi ^{2}}$ which has an opposite sign of the $%
\log $ term compared to the exact series \cite{Rosenstein} $-\frac{a_{T}^{2}%
}{2\beta _{A}}+2\log \frac{\left| a_{T}\right| }{4\pi ^{2}}-\frac{19.9}{%
a_{T}^{2}}$. This is seen on Fig.3 quite clearly. Instead of rising
monotonously from $C/\Delta C=1$ till melting as is predicted by OPT, their
curve (dashed) first drops below $1$ and only later develops a maximum above 
$1.$ In the liquid region it underestimates the specific heat. We conclude
therefore that although the theory of Tesanovic et al is very good at high
temperatures they become of the order $5-10\%$ at $a_{T}=-3$. An advantage
of this theory is that it interpolates smoothly to the solid and never
deviates more than $10\%$.

Experiments on great variety of layered high $T_{c}$ cuprates ($Bi$ or $Tl$ 
\cite{Magn2D} based) show that in 2D, magnetization curves for different
applied fields intersect at a single point $(M^{\ast },T^{\ast })$. The
range of magnetic fields is surprisingly large (from several hundred $Oe$ to
several $Tesla$). This property fixes the scaled LLL magnetization defined
as $m(a_{T})=-\frac{df_{eff}(a_{T})}{da_{T}}=\frac{m_{ab}}{e^{\ast }{h}}%
\sqrt{\frac{4\pi }{b\omega }}M$. Demanding that the first two terms in $%
1/a_{T}^{2}$ expansion of $m(a_{T})$ are consistent with the exact result,
one obtains 
\begin{equation}
m\left( a_{T}\right) =\frac{1}{4}\left( a_{T}-\sqrt{16+a_{T}^{2}}\right)
\label{phen}
\end{equation}
When it is plotted on Fig.2 (the dotted line), we find that at lower
temperatures the magnetization is overestimated. The OPE results are
consistent with the experimental data \cite{Magn2D} (points) within the
precision range till the radius of convergence $a_{T}=-3.$ It is important
to note that deviations of both the phenomenological formula eq.(\ref{phen})
and the Tesanovic's are clearly beyond our error bars. Therefore we conclude
that the coincidence of the intersection of all the lines at the same point $%
(T^{\ast },M^{\ast })$ cannot be exact. Like in 3D the intersection is
approximate, although the approximation is quite good especially at high
magnetic fields.

Specific heat OPE result in 2D is compared on Fig. 3 with Monte Carlo
simulation of the same model by Kato and Nagaosa \cite{MC} (black circles)
(and the phenomenological formula following from eq.(\ref{phen}), dotted
line). The agreement is very good for both the low temperature and the high
temperature OPT.\ 

To summarize, we obtained the optimized perturbation theory results for the
2D and 3D LLL Ginzburg - Landau model in both vortex liquid and solid
phases. The leading approximant (gaussian) gives a rigorous upper bound on
energy, while the convergent series allow one to make several definitive
qualitative conclusions. The intersection of the magnetization lines in only
approximate not only in 3D, but also in 2D. The theory by Tesanovic \cite
{Tesanovic} describes the physics remarkably well at very high temperatures,
but deviates on the 5-10\% precision level at $a_{T}=-2$ in 2D and has
certain imprecise qualitative features in the solid phase. Comparison with
Monte Carlo simulations and some experiments shows excellent agreement.

We are grateful to our colleagues A. Knigavko and T.K. Lee for numerous
discussions and encouragement and Z. Tesanovic for explaining his work to
one of us and sharing his insight.The work was supported by NSC of Taiwan
grant $\#$89-2112-M-009-039.


\begin{figure}[tbp]
\caption{ Optimized (solid lines) and nonoptimized (dashed lines) free
energy approximants in 2D. Numbers indicate order of the approximant.}
\end{figure}

\begin{figure}[tbp]
\caption{ The 2D scaled LLL magnetization. Comparison of data from Jin et al
in ref. $3$ with OPT calculation, Tesanovic et al result of ref. 10 (eq. $9$%
) and phenomenological "interception" theory eq. $10$ are shown for
comparison. }
\end{figure}

\begin{figure}[tbp]
\caption{ Specific heat, 2D. Comparison of MC data with solid OPT (first two
orders), liquid OPT ($10^{th}$ order). Tesanovic et. al. theory and
phenomenological formula are also shown. }
\end{figure}

\end{document}